\documentstyle{article}  
\input{amssym.def} \input{amssym.tex} 
\newtheorem{Definition}{Definition}
\newtheorem{Lemma}{Lemma}
\newtheorem{Theorem}{Theorem}
\newtheorem{Proposition}{Proposition}
\newtheorem{Example}{Example}
\newtheorem{Corollary}{Corollary}

\renewcommand{\ll}{\label}
\newcommand{\be}{\begin{equation}}
\newcommand{\ee}{\end{equation}}
\newcommand{\bea}{\begin{eqnarray}}
\newcommand{\eea}{\end{eqnarray}}
\newcommand{\nn}{\nonumber}
\newcommand{\bib}{\bibitem}
\newcommand{\ci}{\cite}

\newcommand{\ca}{$C^*$-algebra}

\newcommand{\rep}{representation}

\newcommand{\ovl}{\overline}
\newcommand{\wt}{\widetilde}
\newcommand{\til}{\tilde}
\newcommand{\raw}{\rightarrow}

\newcommand{\ot}{\otimes}

\newcommand{\la}{\langle}
\newcommand{\ra}{\rangle}

\newcommand{\rst}{\upharpoonright}
\newcommand{\cov}{\nabla}
\newcommand{\x}{\times}
\newcommand{\hb}{\hbar}

\newcommand{\cin}{C^{\infty}}
\newcommand{\cci}{C^{\infty}_c}
\newcommand{\half}{\mbox{\footnotesize $\frac{1}{2}$}}

\newcommand{\Ah}{{\frak A}^{\hbar}}

\newcommand{\q}{{\cal Q}_{\hbar}}
\newcommand{\lho}{\lim_{\hbar\rightarrow 0}}
\newcommand{\tsr}{T^*\Bbb R^n}

\newcommand{\qw}{{\cal Q}_{\hbar}^W}

\newcommand{\lt}{L^2(\Bbb R^n)}

\newcommand{\inv}{^{-1}}

\newcommand{\Exp}{{\rm Exp}}

\newcommand{\tih}{\times_H}

\newcommand{\daw}{\stackrel{\leftarrow}{\Rightarrow}}

\newcommand{\aaw}{\stackrel{\rightarrow}{\rightarrow} 
\stackrel{\mbox{\tiny $TQ$}}{\mbox{\tiny $Q$}}}
\newcommand{\CPW}{C^{\infty}_{\mbox{\tiny PW}}}

\newcommand{\gm}{\gamma}
\newcommand{\Gm}{\Gamma}

\newcommand{\th}{\theta}

\newcommand{\io}{\iota}
\newcommand{\kp}{\kappa}

\newcommand{\rh}{\rho}

\newcommand{\ta}{\tau}

\newcommand{\phv}{\varphi}

\newcommand{\ps}{\psi}
\newcommand{\Ps}{\Psi}

\newcommand{\A}{{\frak A}}
\newcommand{\B}{{\frak B}}
\newcommand{\GC}{{\frak C}}

\newcommand{\GG}{{\frak G}}
\newcommand{\GI}{{\frak I}}

\newcommand{\g}{{\frak g}}

\newcommand{\CN}{{\cal N}}

\newcommand{\CO}{{\cal O}}

\newcommand{\CQ}{{\cal Q}}

\newcommand{\C}{{\Bbb C}}

\newcommand{\R}{{\Bbb R}}

\newcommand{\SG}{{\sf G}}

\newcommand{\SP}{{\sf P}}

\makeatletter
\newskip\tempskip
\def\endproof{{\parfillskip24\p@ plus\@ne fil\@@par}\tempskip\prevdepth
  \ifdim\lastskip=\z@\tempskip\z@\else\vskip-\lastskip
    \ifdim\tempskip>4\p@ \tempskip.5\tempskip \else \tempskip\z@\fi\fi
  \nobreak\vskip-\baselineskip\vskip-\tempskip\noindent\hbox 
to\hsize{\hfill
    $\blacksquare$}\par\vskip\tempskip\vskip\abovedisplayskip\@doendpe}
\makeatother
\makeatletter
\newskip\tempskip
\def\endiproof{{\parfillskip24\p@ plus\@ne fil\@@par}\tempskip\prevdepth
  \ifdim\lastskip=\z@\tempskip\z@\else\vskip-\lastskip
    \ifdim\tempskip>4\p@ \tempskip.5\tempskip \else \tempskip\z@\fi\fi
  \nobreak\vskip-\baselineskip\vskip-\tempskip\noindent\hbox 
to\hsize{\hfill
    $\Box$}\par\vskip\tempskip\vskip\abovedisplayskip\@doendpe}
\makeatother
\newcommand{\enp}{\endproof}

\renewcommand{\Ah}{{\frak A}_{\hbar}} \newcommand{\Ao}{\tilde{{\frak
A}}_0} \newcommand{\Pri}{\mbox{\rm Prim}} \topmargin = - 0.5 cm
\begin{document} 
\setlength{\baselineskip}{1\baselineskip}
\thispagestyle{empty} \setlength{\unitlength}{1cm} \title{Lie groupoid
$C^*$-algebras and Weyl quantization}
\author{N.P.~Landsman\thanks{Supported by a fellowship from the Royal
Netherlands Academy of Arts and Sciences (KNAW)}\\ \mbox{}\hfill \\
Korteweg-de Vries Institute for Mathematics \\ University of Amsterdam
\\ Plantage Muidergracht 24 \\ 1018 TV AMSTERDAM, THE NETHERLANDS \\
\mbox{}\hfill \\ {\em email:} npl@wins.uva.nl } \date{\today}
\maketitle
\begin{abstract}
A strict quantization of a Poisson manifold $P$ on a subset
$I\subseteq\R$ containing 0 as an accumulation point is defined as a
continuous field of \ca s $\{\Ah\}_{\hbar\in I}$, with $\A_0=C_0(P)$,
a dense subalgebra $\til{\A}_0$ of $C_0(P)$ on which the Poisson
bracket is defined, and a set of continuous cross-sections
$\{\CQ(f)\}_{f\in \til{\A}_0}$ for which $\CQ_0(f)=f$.  Here
$\q(f^*)=\q(f)^*$ for all $\hbar\in I$, whereas for $\hbar\raw 0$ one
requires that $i[\q(f),\q(g)]/\hbar\raw\q(\{f,g\})$ in norm.

For any Lie groupoid $\SG$, the vector bundle $\GG^*$ dual to the
associated Lie algebroid $\GG$ is canonically a Poisson manifold.  Let
$\A_0=C_0(\GG^*)$, and for $\hbar\neq 0$ let $\Ah=C_r^*(\SG)$ be the
reduced \ca\ of $\SG$.  The family of \ca s $\{\Ah\}_{\hbar\in [0,1]}$
forms a continuous field, and we construct a dense subalgebra
$\Ao\subset C_0(\GG^*)$ and an associated family $\{\qw(f)\}$ of
continuous cross-sections of this field, generalizing Weyl
quantization, which define a strict quantization of $\GG^*$.

Many known strict quantizations are a special case of this procedure.
On $P=\tsr$ the maps $\qw(f)$ reduce to standard Weyl quantization;
for $P=T^*Q$, where $Q$ is a Riemannian manifold, one recovers Connes'
tangent groupoid as well as a recent generalization of Weyl's
prescription. When $\SG$ is the gauge groupoid of a principal bundle
one is led to the Weyl quantization of a particle moving in an
external Yang-Mills field. In case that $\SG$ is a Lie group (with Lie
algebra $\g$) one recovers Rieffel's quantization of the Lie-Poisson
structure on $\g^*$.  A transformation group \ca\ defined by a smooth
action of a Lie group on a manifold $Q$ turns out to be the
quantization of the Poisson manifold $\g^*\x Q$ defined by this
action.  \end{abstract}\thispagestyle{empty} \newpage
\section{Introduction}
The notion of quantization to be used in this paper is motivated by
the desire to link the geometric theory of classical mechanics and
reduction \ci{MR,Vai} with the $C^*$-algebraic formulation of quantum
mechanics and induction \ci{MT}, and also with non-commutative
geometry \ci{Con}.  Starting with Rieffel's fundamental paper
\ci{Rie1}, various $C^*$-algebraic definitions of quantization have
been proposed \ci{Rie3,NPL1,Rie4,MT,Rie5}. Definition \ref{defqua}
below is closely related to these proposals, and is particularly
useful in the context of the class of examples studied in this paper.

These examples come from the theory of Lie groupoids and their Lie
algebroids (cf.\ section \ref{LL}). The idea that the \ca\ of a Lie
groupoid is connected to the Poisson manifold defined by the
associated Lie algebroid by (strict) quantization was conjectured in
\ci{NPL1}, and proved in special cases in \ci{NPL2,MT}.
The results of \ci{Rie2,Rie3,NWX} also
supported the claim.  In this paper we prove the conjecture up to
Dirac's condition (\ref{direq}); this is the content of Theorems
\ref{T1} and \ref{T2}.  Following up on our work, Dirac's condition
has finally been proved by Ramazan \ci{Ram}.
This leads to the Corollary at the end of section 5, which is the main
result of the paper.

Further to the examples considered in section \ref{Examples}, it would
be interesting to apply the point of view in this paper to the
holonomy groupoid of a foliation \ci{Con}, and to the Lie groupoid
defined by a manifold with boundary \ci{NWX,Mon}.  Moreover, the
approach to index theory via the tangent groupoid \ci{Con} and its
recent generalization to arbitrary Lie groupoids \ci{MP} may now be
seen from the perspective of `strict' quantization theory. This may be
helpful also in understanding the connection between various other
approaches to index theory which use (formal deformation) quantization
\ci{Fed,ENN2}.

The central notion in $C^*$-algebraic quantization theory is that of a
continuous field of \ca s \ci{Dix}.  For our purposes the following
reformulation is useful \ci{KW}.
\begin{Definition}\ll{defcfca}
A continuous field of \ca s $(\GC,\{\A_x,\phv_x\}_{x\in X})$ over a
locally compact Hausdorff space $X$ consists of a \ca\ $\GC$, a
collection of \ca s $\{\A_x\}_{x\in X}$, and a set
$\{\phv_x:\GC\raw\A_x\}_{x\in X}$ of surjective
$\mbox{}^*$-homomorphisms, such that for all $A\in\GC$
\begin{enumerate}
\item
the function $x\raw \| \phv_x(A)\|$ is in $C_0(X)$;
\item
one has $\|A\| =\sup_{x\in X}\| \phv_x(A)\|$;
\item
there is an element $fA\in\GC$ for any $f\in C_0(X)$ for which
$\phv_x(fA)=f(x)\phv_x(A)$ for all $x\in X$.
\end{enumerate}
\end{Definition}
The continuous cross-sections of the field in the sense of \ci{Dix}
consist of those elements $\{A_x\}_{x\in X}$ of $\prod_{x\in X}\A_x$
for which there is a (necessarily unique)
 $A\in \GC$ such that $A_x=\phv_x(A)$ for all
$x\in X$.

We refer to \ci{MR,Vai} for the theory of Poisson manifolds and
Poisson algebras; the latter is the classical analogue of the
self-adjoint part of a \ca\ \ci{MT}.
\begin{Definition}\ll{defqua}
Let $I\subseteq\R$ contain $0$ as an accumulation point.  A strict
quantization of a Poisson manifold $P$ on $I$ consists of
\begin{enumerate}
\item 
a continuous field of \ca s $(\GC,\{\Ah,\phv_{\hbar}\}_{\hbar\in I})$,
with $\A_0=C_0(P)$;
\item
a dense subspace $\til{\A}_0\subset C_0(P)$ on which the Poisson
bracket is defined, and which is closed under pointwise multiplication
and taking Poisson brackets (in other words, $\til{\A}_0$ is a Poisson
algebra);
\item
a linear map $\CQ:\Ao\raw\GC$ which (with
$\q(f)\equiv\phv_{\hbar}(\CQ(f))$) for all $f\in \til{\A}_0$ and
$\hbar\in I$ satisfies \bea \CQ_0(f)& =& f, \ll{q0f}\\ \q(f^*)& =&
\q(f)^*,\ll{real} \eea and for all $f,g\in \til{\A}_0$ satisfies
Dirac's condition \be \lho \|\frac{i}{\hbar}[\q(f),\q(g)]
-\q(\{f,g\})\|=0. \ll{direq} \ee
\end{enumerate}
\end{Definition}
Elements of $I$ are interpreted as possible values of Planck's
constant $\hbar$, and $\Ah$ is the quantum algebra of observables of
the theory at the given value of $\hbar\neq 0$. For real-valued $f$,
the operator $\q(f)$ is the quantum observable associated to the
classical observable $f$.  This interpretation is possible because of
condition (\ref{real}) in Definition \ref{defqua}.  In view of the
comment after Definition \ref{defcfca}, for fixed $f\in\Ao$ each
family $\{\q(f)\}_{\hbar\in I}$ is a continuous cross-section of the
continuous field in question. In view of (\ref{q0f}) this implies, in
particular, that \be \lho \|\q(f)\q(g)-\q(fg)\| =0. \ll{vneq} \ee This
shows that strict quantization yields asymptotic morphisms in the
sense of $E$-theory \ci{Con}; cf.\ \ci{Nag}.  See \ci{MT} for an
extensive discussion of quantization theory from the above
perspective, including an interpretation of the conditions
(\ref{direq}) and (\ref{vneq}).
\section{Lie groupoids and Lie algebroids\ll{LL}}
Throughout this section, the reader is encouraged to occasionally skip
to section \ref{Examples} to have a look at some examples of the
objects defined.

We refer to \ci{Ren,Mac,CDW,Con,MT,CHW} for the basic definitions on
groupoids; here we merely establish our notation.  Briefly, a groupoid
is a category whose space of arrows $\SG$ is a set (hence the space of
objects $Q$ is a set as well), and whose arrows are all
invertible. The source and target projections are called
$\ta_s:\SG\raw Q$ and $\ta_t:\SG\raw Q$, respectively.

The subset of $\SG\x\SG$ on which the groupoid multiplication (i.e.,
the composition of arrows) is defined is called $\SG_2$; hence
$(\gm_1,\gm_2)\in\SG_2$ iff $\ta_s(\gm_1)=\ta_t(\gm_2)$. The inversion
$\gm\raw\gm\inv$ defines the unit space
$\SG_0=\{\gm\gm\inv|\gm\in\SG\}$, which is related to the base space
$Q$ by the `object inclusion map' $\io:Q\hookrightarrow\SG$; this is a
bijection between $Q$ and $\io(Q)=\SG_0$.  The notation $\SG\daw Q$
for a groupoid to some extent captures the situation.

A Lie groupoid is a groupoid $\SG\daw Q$ where $\SG$ and $Q$ are
manifolds (perhaps with boundary), the maps $\ta_s$ and $\ta_t$ are
surjective submersions, and multiplication and inclusion are smooth
\ci{Mac,CDW,Con,MT,CHW}.  Following \ci{MT}, we now sharpen Def.\
I.2.2 in \ci{Ren}.
\begin{Definition}\ll{deftsystem} A  left Haar system on a Lie
groupoid $\SG\daw Q$ is a family $\{\mu^t_q\}_{q\in Q}$ of positive
measures, where the measure $\mu^t_q$ is defined on $\ta_t\inv(q)$,
such that
\begin{enumerate}
\item
the family is invariant under left-translation in $\SG$;
\item
  each $\mu^t_q$ is locally Lebesgue (i.e., it is equivalent to the
Lebesgue measure in every co-ordinate chart; note that each fiber
$\ta_t\inv(q)$ is a manifold);
\item
for each ${\sf f}\in\cci(\SG)$ the map $q\raw
\int_{\ta_t\inv(q)}d\mu^t_q(\gm) {\sf f}(\gm)$ from $Q$ to $\C$ is
smooth.
\end{enumerate} 
\end{Definition}

Here left-invariance means invariance under all maps $L_{\gm}$,
defined by \be L_{\gm}(\gm'):=\gm\gm' \ll{defLgm} \ee whenever
$(\gm,\gm')\in\SG_2$.  Note that $L_{\gm}$ maps
$\ta_t\inv(\ta_s(\gm))$ diffeomorphically to $\ta_t\inv(\ta_t(\gm))$.

 A Lie groupoid $\SG\daw Q$ has an associated Lie algebroid
\ci{Mac,CDW,MT,CHW}, which we denote by $\GG\aaw$.  This is a vector
bundle over $Q$, which apart from the bundle projection $\ta:\GG\raw
Q$ is equipped with a vector bundle map $\ta_a:\GG\raw TQ$ (called the
anchor), as well as with a Lie bracket $[\, ,\,]_{\GG}$ on the space
$\Gm(\GG)$ of smooth sections of $\GG$, satisfying certain
compatibility conditions.

For our purposes, the essential point in the construction of $\GG\aaw$
from $\SG\daw Q$ lies in the fact that the vector bundle $\GG$ over
$Q$ is the normal bundle $N^{\io}Q$ defined by the embedding
$\io:Q\hookrightarrow \SG$; accordingly, the projection
$\ta:N^{\io}Q\raw Q$ is given by $\ta_s$ or $\ta_t$ (these projections
coincide on $\SG_0$).  The tangent bundle of $\SG$ at the unit space
has a decomposition \be T_{\io(q)}\SG=T_{\io(q)}\SG_0\oplus
T^t_{\io(q)}\SG, \ll{Tioq} \ee where $T^t\SG=\ker(T\ta_t)$ is a
sub-bundle of $T\SG$.  Note that
$T^t_{\gm}\SG=T_{\gm}\ta_t\inv(\ta_t(\gm))$.  Hence $\GG\aaw$ is
isomorphic as a vector bundle to the restriction $\GG'$ of $T^t\SG$ to
$\SG_0$. Under this isomorphism the fiber $\GG_q$ above $q$ is mapped
to the vector space $T^t_{\io(q)}\SG=T_{\io(q)}\ta_t\inv(q)$.

The following pleasant result was pointed out by Ramazan \ci{Ram}.
\begin{Proposition}\ll{razlem}
Every Lie groupoid possesses a left Haar system.
\end{Proposition}

A given strictly positive smooth density $\rh$ on the vector bundle
$\GG$ can be (uniquely) extended to a left-invariant density
$\til{\rh}$ on the vector bundle $T^t\SG$, which in turn yields a left
Haar system by $\mu^t_q(f)=\int_{\ta_t\inv(q)} \til{\rh}f$.  \enp

One may canonically associate a \ca\ $C_r^*(\SG)$ to a Lie groupoid
$\SG\daw Q$ \ci{Con}, and equally canonically associate a Poisson
algebra $\cin(\GG^*)$ to its Lie algebroid $\GG\aaw$ \ci{Cou,CDW}
(here $\GG^*$ is the dual vector bundle of $\GG$, with projection
denoted by $\ta^*$). From the point of view of quantization theory,
these constructions go hand in hand \ci{NPL1,NPL2,MT}.

Although a left Haar system is not intrinsic, and an intrinsic
definition of $C_r^*(\SG)$ may be given \ci{Con,MT,Ram}, it vastly
simplifies the presentation of our results if we define this \ca\
relative to a particular choice of a left Haar system
$\{\mu^t_q\}_{q\in Q}$. For ${\sf f},{\sf g}\in\cci(\SG)$ the product
$*$ in $C_r^*(\SG)$ is then given by the convolution \ci{Ren} \be {\sf
f}*{\sf g}(\gm):=\int_{\ta_t\inv(\ta_s(\gm))}
d\mu^t_{\ta_s(\gm)}(\gm_1)\, {\sf f}(\gm \gm_1){\sf g}(\gm_1\inv);
\ll{oldconv} \ee the involution is defined by \be {\sf
f}^*(\gm):=\ovl{{\sf f}(\gm\inv)}.\ll{oldinv} \ee The reduced groupoid
\ca\ $C_r^*(\SG)$ is the completion of $\cci(\SG)$ in a suitable
$C^*$-norm \ci{Con,Ren,MT}.

On the classical side, the Poisson algebra $\cin(\GG^*)$ associated to
a Lie algebroid $\GG$ \ci{Cou,CDW,MT} is most simply defined by
listing special cases which uniquely determine the Poisson bracket.
These are
\bea \{f,g\} & = & 0; \ll{pblieoid1} \\
\{\til{s},f\} & = & - \ta_a\circ s f; \ll{pblieoid2} \\
\{\til{s}_1,\til{s}_2\} & = & -
\wt{[s_1,s_2]_{\GG}}. \ll{pblieoid3} \eea Here $f,g\in\cin(Q)$
(regarded as functions on $\GG^*$ in the obvious way), and
$\til{s}\in\cin(\GG^*)$ is defined by a section $s$ of $\GG$ through
$\til{s}(\th)= \th(s(\ta^*(\th)))$, etc.  See \ci{CDW} for an
intrinsic definition. 
\section{A generalized  exponential map\ll{ExpWeyl}}
Throughout the remainder of the paper, $\GG\aaw$ will be the Lie
algebroid of a Lie groupoid $\SG\daw Q$.  In order to state and prove
our main results we need to construct an exponential map
$\Exp^W:\GG\raw\SG$, which generalizes the map $\Exp$ from a Lie
algebra to an associated Lie group. The construction of such a map was
outlined by Pradines \ci{Pra}, but in order to eventually satisfy the
self-adjointness condition (\ref{real}) on our quantization map we
need a different construction \ci{MT}. As in \ci{Pra}, our exponential
map depends on the choice of a connection on the vector bundle $\GG$.
As before, the reader is referred to section \ref{Examples} for
examples of the constructions below.  \begin{Lemma}\ll{pradexp} The
vector bundles $T^t\SG$ and $\ta_s^*\GG$ (over $\SG$) are isomorphic.
\end{Lemma}

 The pull-back bundle $\ta_s^*\GG$ is a vector bundle over $\SG$ with
projection onto the second variable.  The isomorphism is proved via
the vector bundle isomorphism $\GG\simeq\GG'$; see section \ref{LL}.
Recalling (\ref{defLgm}), one checks that $TL_{\gm\inv}:
T^t_{\gm}\SG\raw T^t_{\gm\inv\gm}\SG$ is the desired bundle
isomorphism between $T^t\SG$ and $\ta_s^*\GG'$.  \enp

Let us now assume that $\GG$ has a covariant derivative (or,
 equivalently, a connection), with associated horizontal lift
 $\ell^{\GG}$.  By Lemma \ref{pradexp} one then obtains a connection
 on $T^t\SG$ (seen as a vector bundle over $\SG$, whose projection is
 borrowed from $T\SG$) through pull-back.  Going through the
 definitions, one finds that the associated horizontal lift $\ell$ of
 a tangent vector $X=\dot{\gm}:=d\gm(t)/dt_{t=0}$ in $T_{\gm}\SG$ to
 $Y\in T_{\gm}^t\SG$ is \be \ell_Y(\dot{\gm})=\frac{d}{dt}
 [L_{\gm(t)*}
 \ell^{\GG}_{TL_{\gm\inv}Y}(\ta_s(\gm(t)))]_{t=0},\ll{assconncom} \ee
 which is an element of $T_Y(T^t\SG)$ (here $\ell^{\GG}(\ldots)$ lifts
 a curve).

Since the bundle $T^t\SG\raw\SG$ has a connection, one can define
geodesic flow $X\raw X(t)$ on $T^t\SG$ in precisely the same way as on
a tangent bundle with affine connection. That is, the flow $X(t)$ is
the solution of \be \dot{X}(t)=\ell_{X(t)}(X(t)), \ll{dotXtis} \ee
with initial condition $X(0)=X$.
\begin{Definition}\ll{defExpL}
Let the Lie algebroid $\GG\aaw$ of a Lie groupoid $\SG\daw Q$ be
equipped with a connection. Relative to the latter, the left
exponential map $\Exp^L:\GG\raw\SG$ is defined by \be
\Exp^L(X):=\gm_{X'}(1)=\ta_{T^t\SG\raw \SG}(X'(1)), \ll{ExpL} \ee
whenever the geodesic flow $X'(t)$ on $T^t\SG$ (defined by the
connection on $T^t\SG$ pulled back from the one on $\GG$) is defined
at $t=1$. Here $X'\in\GG'=T^t\SG\rst\SG_0$ is the image of $X$ under
the isomorphism $\GG'\simeq \GG$. \end{Definition}

Our goal, however, is to define a `symmetrized' version of $\Exp^L$.
\begin{Lemma}
For all $X\in\GG$ for which $\Exp^L(X)$ is defined one has \be
\ta_t(\Exp^L(X))=\ta(X).\ll{tatExpL} \ee \end{Lemma}

Here $\ta$ is the bundle projection of the Lie algebroid.  We write
$X$ for $X'$ in (\ref{ExpL}). One has $\ta_t(\gm_{X}(0))=\ta(X)$ and
$$
\frac{d}{dt} \ta_t(\gm_{X}(t))=T(\ta_t\circ\ta_{T^t\SG\raw
\SG})\ell_{X(t)}(X(t))=T\ta_t X(t)=0,
$$
since $\ell_X(Y)$ covers $Y$, and $X(t)\in T^t\SG=\ker(T\ta_t)\cap
T\SG$.  \enp

We combine this with the obvious $\ta(\half X)=\ta(-\half X)$ to infer
that
$$\ta_t(\Exp^L(\half X))=\ta_t(\Exp^L(-\half X))=\ta_s(\Exp^L(-\half
X)\inv).$$ Thus the (groupoid) multiplication in (\ref{ExpW}) below is
well-defined.
\begin{Definition}\ll{defExpW}
The Weyl exponential map $\Exp^W:\GG\raw\SG$ is defined by \be
\Exp^W(X):=\Exp^L(-\half X)\inv \Exp^L(\half X).\ll{ExpW} \ee
\end{Definition}
 
The following result is closely related to the tubular neighbourhood
theorem.
\begin{Proposition}\ll{tntoid}
The maps $\Exp^L$ and $\Exp^W$ are diffeomorphisms from a
neighbourhood $\CN^{\io}$ of $Q\subset\GG$ (as the zero section) to a
neighbourhood $\CN_{\io}$ of $\io(Q)$ in $\SG$, such that
$\Exp^L(q)=\Exp^W(q)=\io(q)$ for all $q\in Q$.
\end{Proposition}

The property $\Exp^L(q)= \io(q)$ is immediate from Definition
\ref{defExpL}.  The push-forward of $\Exp^L$ at $q$ is
$T\Exp^L:T_q\GG\raw T_{\io(q)}\SG$.  Now recall the decomposition
(\ref{Tioq}). For $X$ tangent to $Q\subset\GG$ one immediately sees
that $T\Exp^L(X)=T\io(X)$.  For $X$ tangent to the fiber $\ta\inv(q)$,
which we identify with $T_{\io(q)}^t\SG$, one has $T\Exp^L(X)=X'$, as
follows by the standard argument used to prove that $\exp_q$ in the
theory of affine geodesics is a local diffeomorphism: for a curve
$X(s)=sX$ in $T_{\io(q)}^t\SG$ one has $\Exp^L(X(s))=\gm_{X'(s)}(1)
=\gm_{X'}(s)$, so that $d/ds[ \Exp^L(X(s))]_{s=0}=X'$.

Since $T\Exp^L$ is a bijection at $q$, the inverse function theorem
implies that $\Exp^L$ is a local diffeomorphism. Since it maps $Q$
pointwise to $\io(Q)$, the local diffeomorphisms can be patched
together to yield a diffeomorphism of the neighbourhoods stated in
Proposition \ref{tntoid}; we omit the details of this last step, since
it is identical to the proof of the tubular neighbourhood theorem.

As for $\Exp^W$, for $X\in T_qQ\subset T_q\GG$ we have
$T\Exp^W(X)=T\io(X)$.  Also, $$ \frac{d}{ds} [\Exp^L(-\half sX)\inv
\Exp^L(\half sX)]_{s=0}= -\half TI(X')+\half X',
$$
where $TI$ is the push-forward of the inversion $I$ in $\SG$. The
right-hand side lies in $\ker(T\ta_s+T\ta_t)\subset T\SG$, and every
element in this kernel is of the stated form.  Similarly to
(\ref{Tioq}), one may prove the decomposition \be
T_{\io(q)}\SG=T_{\io(q)}\SG_0\oplus
\ker(T\ta_s+T\ta_t)(\io(q)). \ll{Tioq2} \ee It follows that $T\Exp^W$
is a bijection at $q$, and the second part of the theorem is derived
as for $\Exp^L$.\enp
\section{The normal groupoid and continuous fields of \ca s} 
We now come to the first part of the proof of the conjecture that
 $C^*_r(\SG)$ is related to the Poisson manifold $\GG^*$ by a strict
 quantization.
\begin{Theorem}\ll{T1}
Let $\SG$ be a Lie groupoid, with associated Lie algebroid $\GG$.
Take $I=[0,1]$ and put $\A_0=C_0(\GG^*)$, where $\GG^*$ is the dual
vector bundle of $\GG$, and $\Ah=C^*_r(\SG)$ for $\hbar\in I\backslash
\{0\}$.

There exists a \ca\ $\GC$ and a family of surjective
$\mbox{}^*$-homomorphisms
$\{\phv_{\hbar}:\GC\raw\A_{\hbar}\}_{\hbar\in I}$ such that
$(\GC,\{\Ah,\phv_{\hbar}\}_{\hbar\in I})$ is a continuous field of \ca
s.
\end{Theorem}

The proof uses the normal groupoid of Hilsum and Skandalis \ci{HS}
(also cf.\ \ci{Wei1,MT}), re-interpreted in terms of the Lie
algebroid. We recall the definition; our construction of the smooth
structure is different from the one in \ci{HS}.  The essence is to
regard the vector bundle $\GG$ as a Lie groupoid under addition in
each fiber, and glue it to $\SG$ so as to obtain a new Lie groupoid
containing both $\SG$ and $\GG$.
\begin{Definition}\ll{defnormoid} 
Let $\SG\daw Q$ be a Lie groupoid with associated Lie algebroid
$\GG\aaw$.  The normal groupoid $\SG_N$ is a Lie groupoid with base
$[0,1]\x Q$, defined by the following structures.
\begin{itemize}
\item
As a set, $\SG_N=\GG\cup \{(0,1]\x\SG\}$. We write elements of $\SG_N$
as pairs $(\hbar,u)$, where $u\in\GG$ for $\hbar=0$ and $u\in\SG$ for
$\hbar\neq 0$.  Thus $\GG$ is identified with $\{0\}\x \GG$.  \item As
a groupoid, $\SG_N=\{0\x\GG\} \cup \{ (0,1]\x\SG\}$.  Here $\GG$ is
regarded as a Lie groupoid over $Q$, with $\ta_s=\ta_t=\ta$ and
addition in the fibers as the groupoid multiplication.  The groupoid
operations in $(0,1]\x\SG$ are those in $\SG$.
\item
The smooth structure on $\SG_N$, making it a manifold with boundary,
is as follows.  To start, the open subset $\CO_1:=(0,1]\x\SG\subset
\SG_N$ inherits the product manifold structure.  Let $Q\subset
\CN^{\io}\subset\GG$ and $\io(Q)\subset\CN_{\io}\subset\SG$, as in
Theorem \ref{tntoid}. Let $\CO$ be the open subset of $[0,1]\x \GG$
(equipped with the product manifold structure; this is a manifold with
boundary, since $[0,1]$ is), defined as $\CO:=\{(\hbar,X)\,|\,\hbar
X\in \CN^{\io}\}$.  Note that $\{0\}\x\GG\subset \CO$.  The map
$\rh:\CO\raw \SG_N$ is defined by \bea \rh(0,X) & := & (0,X); 
\ll{OX}\nn \\
\rh(\hbar,X) & := & (\hbar, \Exp^W(\hbar X)). 
\ll{Whbar} \eea Since $\Exp^W:
\CN^{\io}\raw \CN_{\io}$ is a diffeomorphism (cf.\ Proposition
\ref{tntoid}) we see that $\rh$ is a bijection from $\CO$ to
$\CO_2:=\{0\x\GG\} \cup \{(0,1]\x\CN_{\io}\}$.  This defines the
smooth structure on $\CO_2$ in terms of the smooth structure on $\CO$.
Since $\CO_1$ and $\CO_2$ cover $\SG_N$, this specifies the smooth
structure on $\SG_N$.
\end{itemize} 
\end{Definition}

The fact that $\SG_N$ is a Lie groupoid eventually follows from the
 corresponding property of $\SG$.  The given chart is defined in terms
 of the Weyl exponential, which depends on the choice of a connection
 in $\GG$. However, one may verify that any (smooth) connection, or,
 indeed, any ($Q$-preserving) diffeomorphisms between $\CN^{\io}$ and
 $\CN_{\io}$ leads to an equivalent smooth structure on $\SG_N$.  For
 example, we could have used $\Exp^L$ instead of $\Exp^W$.  Also, the
 smoothness of $\Exp^W$ makes the above manifold structure on $\SG_N$
 well defined, in that open subsets of $\CO_1\cap\CO_2$ are assigned
 the same smooth structure.

Since $\SG_N$ is a Lie groupoid, we can form the \ca\ $C^*_r(\SG_N)$,
which plays the role of $\GC$ in Theorem \ref{T1}. To proceed, we need
a result due to Lee \ci{Lee}.
\begin{Lemma}\ll{tomiyama}
Let $\GC$ be a \ca, and let $\ps:\Pri(\GC)\raw X$ be a continuous and
open map from the primitive spectrum $\Pri(B)$ (equipped with the
Jacobson topology \ci{Dix}) to a locally compact Hausdorff space $X$.
Define $\GI_{x}:=\cap \ps\inv(x)$; i.e., $A\in\GI_{x}$ iff
$\pi_{\GI}(A)=0$ for all $\GI\in\ps\inv(x)$ (here $\pi_{\GI}(\GC)$ is
the irreducible \rep\ whose kernel is $\GI$).  Note that $\GI_x$ is a
(closed two-sided) ideal in $\GC$.

Taking $\A_x= \GC/\GI_x$ and $\phv_x:\GC\raw\A_x$ to be the canonical
projection, $(\GC,\{\A_x,\phv_x\}_{x\in X})$ is a continuous field of
\ca s.
\end{Lemma}

For the proof cf.\ \ci{ENN1}. We apply this lemma with
$\GC=C_r^*(\SG_N)$ and $X=I=[0,1]$. In order to verify the assumption
in the lemma, we first note that $\GI_0\simeq C_0((0,1])\ot
C^*_r(\SG)$, as follows from a glance at the topology of
$\SG_N$. Hence $\Pri(\GI_0)=(0,1]\x \Pri(C^*_r(\SG))$, with the
product topology.  Furthermore, one has $C_r^*(\SG_N)/\GI_0\simeq
C_r^*(\GG)\simeq C_0(\GG^*)$; the second isomorphism is established by
the fiberwise Fourier transform (\ref{fouroidinv}) below (also cf.\
\ci{HS,Con}).  Hence $\Pri(C^*(\SG_N)/\GI_0)\simeq\GG^*$.  Using this
in Prop.\ 3.2.1 in \ci{Dix}, with $A=C^*_r(\SG_N)$ and $I$ the ideal
$\GI_0$ generated by those $f\in\cci(\SG_N)$ which vanish at
$\hbar=0$, yields the decomposition \be \Pri(C^*_r(\SG_N))\simeq
\GG^*\cup \{(0,1]\x \Pri(C^*_r(\SG))\}, \ll{decspgroid} \ee in which
$\GG^*$ is closed.  This does not provide the full topology on
$\Pri(C^*_r(\SG_N))$, but it is sufficient to know that $\GG^*$ is not
open. If it were, $(0,1]\x \Pri(C^*_r(\SG))$ would be closed in
$\Pri(C^*_r(\SG_N))$, and this possibility can be safely be excluded
by looking at the topology of $\SG_N$ and the definition of the
Jacobson topology.

Using (\ref{decspgroid}), we can define a map $\ps:\Pri(C^*_r(\SG_N))
\raw [0,1]$ by $\ps(\GI)=0$ for all $\GI\in\GG^*$ and
$\ps(\hbar,\GI)=\hbar$ for $\hbar\neq 0$ and $\GI\in
\Pri(C^*_r(\SG))$.  It is clear from the preceding considerations that
$\ps$ is continuous and open. Using this in Lemma \ref{tomiyama}, one
sees that $\GI_{\hbar}$ is the ideal in $C^*_r(\SG_N)$ generated by
those $f\in\cci(\SG_P)$ which vanish at $\hbar$. Hence $\A_0\simeq
C_0(\GG^*)$, as above, and $\Ah\simeq C^*_r(\SG)$ for $\hbar\neq
0$. Theorem \ref{T1} then follows from Lemma \ref{tomiyama}.  \enp

As pointed out to the author by G. Skandalis (private communication,
June 1997), similar considerations lead to the following
generalization of Theorem \ref{T1}.
 
Let $\til{\SG}$ be a Lie groupoid with base $\til{Q}$, and let $p$ be
a continuous and open map from $\til{Q}$ to some Hausdorff space $X$,
which is $\til{\SG}$-invariant in the sense that
$p\circ\ta_s=p\circ\ta_t$. Define $\til{\SG}_x:=(p\circ\ta_s)\inv(x)$
(this is a sub-groupoid of $\til{\SG}$ because of the
$\til{\SG}$-invariance of $p$), and $\A^x:=C^*(\til{\SG}_x)$.  Then
the collection $(\{\A^x\}_{x\in X},C^*(\til{G}))$ is a continuous
field of \ca s at those points $x$ where
$C^*(\til{\SG}_x)=C_r^*(\til{\SG}_x)$. Here $f\in C^*(\til{G})$ is
understood to define a section of the field $\{\A^x\}_{x\in X}$ by
$f(x)=f\rst \til{\SG}_x$.

We apply this to our situation by taking $\til{\SG}=\SG_N$ and $X=I$,
 hence $\til{Q}=I\x Q$, and $p$ is just projection onto the first
 variable. Continuity away from $\hbar=0$ follows from the triviality
 of the field for $\hbar\neq 0$ (whether or not
 $C^*_r(\SG)=C^*(\SG)$).  Continuity at $\hbar =0$ follows by noticing
 that $C^*_r(\GG)=C^*(\GG)$, both sides are equal to $C_0(\GG^*)$.  In
 other words, from this point of view it is the amenability of $\GG$,
 regarded as a Lie groupoid, that lies behind Theorem \ref{T1}.
\section{Weyl quantization on the dual of a Lie algebroid\ll{WQ}}
Let $\GG\aaw$ be a Lie algebroid, with bundle projection $\ta$.  We
start by defining a fiberwise Fourier transform
$\grave{f}\in\cin(\GG)$ of suitable $f\in\cin(\GG^*)$. This transform
depends on the choice of a family $\{\mu^L_q\}_{q\in Q}$ of Lebesgue
measures, where $\mu^L_q$ is defined on the fiber $\ta\inv(q)$. We
will discuss the normalization of each $\mu^L_q$ in the proof of
Theorem \ref{T2}; for the moment we merely assume that the
$q$-dependence is smooth in the obvious (weak) sense.  For a function
$\grave{f}$ on $\GG$ which is $L^1$ on each fiber we put \be
f(\th)=\int_{\ta\inv(q)} d\mu^L_q(X)\, e^{-i\th(X)}\grave{f}(X),
\ll{fouroidinv} \ee where $X\in\ta\inv(q)$.  Each $\mu^L_q$ determines
a Lebesgue measure $\mu^{L*}_q$ on the fiber $\ta_{\GG^*\raw Q}\inv
(q)$ of $\GG^*$ by fixing the normalization in requiring that the
inverse to (\ref{fouroidinv}) is given by \be
\grave{f}(X)=\int_{\ta_{\GG^*\raw Q}\inv (q)} d\mu^{L*}_q (\th)\,
e^{i\th(X)}f(\th).\ll{fouroid} \ee
 
Having constructed a Fourier transform, we define the class
$\CPW(\GG^*)$ as consisting of those smooth functions on $\GG^*$ whose
Fourier transform is in $\cci(\GG)$; this generalizes the class of
Paley-Wiener functions on $\tsr\simeq \C^n$.  We pick a function
$\kp\in\cin(\GG,\R)$ with support in $\CN^{\io}$ (cf.\ Proposition
\ref{tntoid}), equalling unity in some smaller tubular neighbourhood
of $Q$, as well as satisfying $\kp(-X)=\kp(X)$ for all $X\in\GG$.
\begin{Definition}\ll{Weylultimate}
Let $\SG$ be a Lie groupoid with Lie algebroid $\GG$.  For $\hbar\neq
 0$, the  Weyl quantization of $f\in C^{\infty}_{\mbox{\tiny
 PW}}(\GG^*)$ is the element $\qw(f)\in\cci(\SG)$, regarded as a
 dense subalgebra of $C^*_r(\SG)$, defined by $\qw(f)(\gm):=0$
 when $\gm\notin \CN_{\io}$, and by \be
 \qw(f)(\Exp^W(X)):=\hbar^{-n}\kp(X)\grave{f}(
 X/\hbar). \ll{qwfEW} \ee Here the Weyl exponential
 $\Exp^W:\GG\raw\SG$ is defined in (\ref{ExpW}), and the cutoff
 function $\kp$ is as specified above.\end{Definition}

This definition is possible by virtue of Proposition \ref{tntoid}. By
 our choice of $\CPW(\GG^*)$, the operator $\qw(f)$ is
 independent of $\kp$ for small enough $\hbar$ (depending on $f$).
\begin{Theorem}\ll{T2}
Let $\SG$ be a Lie groupoid with Lie algebroid $\GG\aaw$, and take
$\Ao=\CPW(\GG^*)$.  For each $f\in\Ao$ operator $\qw(f)$ of
Definition \ref{Weylultimate} satisfies $\qw(f)^*=\qw(f^*)$, and the
family $\{\qw(f)\}_{\hbar\in [0,1]}$, with $\CQ^W_0(f)=f$, is a
continuous cross-section of the continuous field of \ca s of Theorem
\ref{T1}.
\end{Theorem}

Writing the Poisson bracket and the pointwise product in terms of the
Fourier transform, one quickly establishes that $\Ao$ is indeed a
Poisson algebra.

It is immediate from (\ref{oldinv}) and (\ref{ExpW}) that for
real-valued $f\in\Ao$ the operator $\qw(f)$ is self-adjoint in
$C^*_{(r)}(\SG)$; this implies the first claim.

To prove the second claim, we pick a left Haar system
$\{\mu^t_q\}_{q\in Q}$ on $\SG\daw Q$; see Proposition
\ref{razlem}. The vector bundle $\GG$, regarded as a Lie groupoid
under addition in each fiber (cf.\ Definition \ref{defnormoid}), has a
left Haar system in any case, consisting of the family
$\{\mu^L_q\}_{q\in Q}$ of Lebesgue measures on each fiber already used
in the construction of the Fourier transform.  Since we have a Lie
groupoid, the Radon-Nikodym derivative
$J_q(X):=d\mu^t_q(\Exp^W(X))/d\mu^L_q(X)$ is well defined and strictly
positive on $\CN^{\io}$ (since both measures are locally Lebesgue on
spaces with the same dimension). We now fix the normalization of the
$\mu^L_q$ by requiring that $\lim_{X\raw 0} J_q(X)=1$ for all $q$.
This leads to a left Haar system for $\SG_N$, given by \bea
\mu^t_{(0,q)} & := & \mu^L_q; \nn \\ \mu^t_{(\hbar,q)} & := &
\hbar^{-n}\mu^t_q, \ll{tploid} \eea where $n$ is the dimension of the
typical fiber of $\GG$. The factor $\hbar^{-n}$ is necessary in order
to satisfy condition 3 in Definition \ref{deftsystem} at $\hbar=0$, as
is easily verified using the manifold structure on $\SG_N$.

Thus the $\mbox{}^*$-algebraic structure on $\cci(\SG_N)$ defined by
(\ref{oldconv}) and (\ref{oldinv}) with \ref{defnormoid} and
(\ref{tploid}) becomes \bea {\sf f}*{\sf g}(0,X) & = &
\int_{\ta\inv\circ\ta(X)}d\mu_{\ta(X)}^L(Y)\, {\sf f}(0,X-Y){\sf
g}(0,Y); \\ {\sf f}*{\sf g}(\hb,\gm) & = & \hbar^{-n}
\int_{\ta_t\inv(\ta_s(\gm))} d\mu^t_{\ta_s(\gm)}(\gm_1)\, {\sf
f}(\hbar,\gm\gm_1){\sf g}(\hbar,\gm_1\inv) ; \ll{eq1} \\ {\sf
f}^*(0,X) & = & \ovl{{\sf f}(0,-X)} ; \\ {\sf f}^*(\hb,\gm) & = &
\ovl{{\sf f}(\hb,\gm\inv)}.  \ll{longea} \eea

One sees that, for given $f\in\CPW(\GG^*)$, the function $\CQ(f)$ on
$\SG_N$ defined by $\CQ(f)(0,X)=\grave{f}(X)$,
$\CQ(f)(\hbar,\Exp^W(X))= \kp(X)\grave{f}(X/\hbar)$, and
$\CQ(f)(\hbar,\gm)= 0$ for $\gm\notin\CN_{\io}$, is smooth on $\SG_N$;
cf.\ Definition \ref{defnormoid}. In other words, $\CQ(f)$ is an
element of $C^*_r(\SG_N)$.  

Recall that $\GI_{\hbar}$ is the ideal in $C^*_r(\SG_N)$ generated by
those functions in $\cci(\SG_N)$ which vanish at $\hbar$. The
canonical map ${\sf f}\raw [{\sf f}]_{\hbar}$ from $C^*_r(\SG_N)$ to
$C^*_r(\SG_N)/\GI_{\hbar}$ is given, for $\hbar\neq 0$, by $[{\sf
f}]_{\hbar}(\cdot)={\sf f}(\hbar,\cdot)$. However, in view of the
factor $\hbar^{-n}$ in (\ref{eq1}), this map is only a
$\mbox{}^*$-homomorphism from $C^*_r(\SG_N)$ to $C^*_r(\SG)$ if we add
a factor $\hbar^{-n}$ to the definition (\ref{oldconv}) of convolution
on $\SG$.  Since for $\hbar\neq 0$ we would like to identify
$C^*_r(\SG_N)/\GI_{\hbar}$ with $C^*_r(\SG)$, in which convolution is
defined in the usual, $\hbar$-independent way, we should therefore
define the maps $\phv_{\hbar}$ of Theorem
\ref{T1} by \bea \phv_0({\sf f}) : \th & \mapsto & \acute{\sf f}(0,\th);
\nn \\ \phv_{\hbar}({\sf f}) : \gm & \mapsto & \hbar^{-n}{\sf
f}(\hbar,\gm) \:\:\: (\hbar\neq 0). \ll{phvhb} \eea Here 
$\phv_0:C^*_r(\SG_N)\raw C_0(\GG^*)$, and
$\acute{\sf f}(0,\th)$ and ${\sf
f}(0,X)$ are related as $f(\th)$ and $\grave{f}(X)$ are in
(\ref{fouroidinv}).  For $\hbar\neq 0$ one of course has 
$\phv_{\hbar}:C^*_r(\SG_N)\raw C^*_r(\SG)$.
These expressions are initially defined for ${\sf
f}\in\cci(\SG_N)$; since $\phv_{\hbar}$ is contractive, they are
subsequently extended to general ${\sf f}\in C^*_r(\SG_N)$ by
continuity.
  
This explains the factor $\hbar^{-n}$ in (\ref{qwfEW}); the theorem
then follows from the paragraph after (\ref{longea}).\enp

The important calculations of Ramazan \ci{Ram} show that \be
\lim_{\hbar\raw
0}\|
\frac{i}{\hb}[\qw(f),\qw(g)]-\qw(\{f,g\})\|
=0.\ll{DirL}
\ee for all $f,g\in\Ao$; this is Dirac's condition 
(he in addition proves this to hold in  formal
deformation quantization). 
\begin{Corollary}
Let $\SG$ be a Lie groupoid, with associated
\begin{itemize}
\item
Lie algebroid $\GG\aaw$;
\item
 Poisson manifold
 $\GG^*$ (the dual bundle to $\GG$, with Poisson structure
(\ref{pblieoid1})--(\ref{pblieoid3}));
\item
normal groupoid $\SG_N$ (cf.\ Definition
\ref{defnormoid}).
\end{itemize}
In the context of Definition \ref{defqua}, the ingredients listed below
yield a strict quantization of the Poisson manifold $P=\GG^*$:
\begin{enumerate}
\item
The continuous field of \ca s given by
$\GC=C_r^*(\SG_N)$,  $\A_0=C_0(\GG^*)$, 
  $\Ah=C^*_r(\SG)$ for $\hbar\in I\backslash
\{0\}$, and $\phv_{\hbar}$ as defined in (\ref{phvhb}); cf.\ Theorem \ref{T1}.
\item
  The dense subspace  $\Ao=\CPW(\GG^*)$ of fiberwise Paley--Wiener 
functions on $\GG^*$ (as
defined below (\ref{fouroid})).
\item
  The map $\CQ:\CPW(\GG^*)\raw C_r^*(\SG_N)$  defined by putting
$\CQ_{\hbar}=\qw$ (as specified in Definition \ref{Weylultimate});
this determines $\CQ$ by Theorem \ref{T2} and the remark after
Definition \ref{defcfca}.
\end{enumerate}
\end{Corollary}
\section{Examples\ll{Examples}}
In this section we illustrate the concepts introduced above, and show
that a number of known strict quantizations are special cases of
Corollary 1. Details of these examples will be
omitted; see \ci{Mac,CDW,MT,CHW} for matters related to the Lie
groupoids and Lie algebroids involved, and cf.\ \ci{Con,Ren,MT,Ram}
for the \ca s that appear. The quantization maps are discussed in
detail in \ci{MT}.

It turns out that a number of examples are more naturally described
by changing some signs, as follows. We denote $\GG^*$, seen as a
Poisson manifold through (\ref{pblieoid1})--(\ref{pblieoid3}), by
$\GG^*_+$. Alternatively, we may insert minus signs on the right-hand
sides of (\ref{pblieoid2}) and (\ref{pblieoid3}), defining the
Poisson manifold $\GG^*_-$.
The normal groupoid $\SG_N$ may be equipped with a different 
manifold structure 
 by replacing $\Exp^W(\hbar X)$ in (\ref{Whbar}) by $\Exp^W(-\hbar X)$;
the original Definition \ref{defnormoid} yields a manifold $\SG_N^+$, and
the modified one defines $\SG_N^-$.
 (The original  smooth structure is equivalent to
the modified  one by the diffeomorphism $(0,X)\mapsto (0,-X)$ and
$(\hbar,\gm)\mapsto(\hbar,\gm)$.)
In (\ref{qwfEW}) we may replace $\grave{f}(X/\hbar)$ by
$\grave{f}(-X/\hbar)$, defining a quantization map
$\qw(\cdot)_-$, differing from the original one $\qw(\cdot)_+=\qw(\cdot)$.

Theorems \ref{T1} and \ref{T2}, eq.\ (\ref{DirL}),
 as well as  Corollary 1 remain valid
if all signs are simultaneously changed in this way.
\begin{Example}\ll{WM} 
Weyl quantization on a manifold.
\end{Example}

The pair groupoid $Q\x Q\daw Q$ on a set $Q$ is defined by the
operations $\ta_s(q_1,q_2):=q_2$, $\ta_t(q_1,q_2):=q_1$,
$\io(q):=(q,q)$, $(q_1,q_2)\cdot (q_2,q_3):=(q_1,q_3)$, and
$(q_1,q_2)\inv:=(q_2,q_1)$.  This is a Lie groupoid when $Q$ is a
manifold.  Any measure $\nu$ on $Q$ which is locally Lebesgue defines
a left Haar system. One has $C_r^*(Q\x Q)\simeq\B_0(L^2(Q))$, the \ca\
of all compact operators on $L^2(Q,\nu)$.

The associated Lie algebroid is the tangent bundle $TQ$, with the
usual bundle projection and Lie bracket, and the anchor is the
identity.  The Poisson bracket on $T^*Q$ is the canonical one.

To define $\Exp^W$ one chooses an affine connection $\cov$ on $TQ$,
with associated exponential map $\exp:TQ\raw Q$. Then \bea \Exp^L(X) &
= & (\ta(X),\exp_{\ta(X)}(X)) ;\ll{expLQQ} \\ \Exp^W(X) & = &
(\exp_{\ta(X)}(-\half X),\exp_{\ta(X)}(\half X)),\ll{expWQQ} \eea
where $X\in TQ$ and $\ta:=\ta_{TQ\raw Q}$.

 On $Q=\R^n$ with flat metric and corresponding flat Riemannian
connection this simplifies to $\Exp^W(v,q)=(q-\half v,q+\half v)$,
where we have used canonical co-ordinates on $T\R^n$.  The operator
$\qw(f)_-$ on $\lt$ defined by (\ref{qwfEW}), where one may take
$\kp=1$, with (\ref{fouroid}), is then given by \be
\qw(f)_-\Ps(x)=\int_{T^*\R^n} \frac{d^npd^ny}{(2\pi\hbar)^n}\,
e^{ip(x-y)/\hbar}f(p,\half(x+y) )\Ps(y). \ll{defweylq} \ee This is
Weyl's original prescription. The associated continuous field of \ca s
is $\A_0=C_0(\tsr)$ and $\A_{\hbar}=\B_0(\lt)$ for $\hbar\neq 0$.  The
fact that this quantization map is strict, and in particular satisfies
(\ref{direq}), was proved by Rieffel \ci{Rie3}; also cf.\
\ci{MT}. Replacing $I=[0,1]$, as we have used so far in connection
with Definition \ref{defqua}, by $I=\R$, the \ca\ $\GC$ in Definition
\ref{defcfca} is $C_r^*(H_n)=C^*(H_n)$, the (reduced) group algebra of
the simply connected Heisenberg group on $\R^n$ \ci{ENN1}. This is
indeed the reduced \ca\ of the tangent groupoid of $\R^n$ (see below).

When $Q$ is an arbitrary manifold, the normal groupoid $(Q\x Q)_N$ is
 the tangent groupoid of $Q$ \ci{Con}.  If one takes the affine
 connection on $TQ$ to be the Levi-Civita connection given by a
 Riemannian metric on $Q$, one recovers the extension of Weyl's
 prescription considered in \ci{NPL1,MT}. One now has $\A_0=C_0(T^*Q)$
 and $\A_{\hbar}=\B_0(L^2(Q))$ for $\hbar\neq 0$, and $\qw$ duly
 satisfies (\ref{direq}); see \ci{NPL1,MT}, where references to
 alternative generalizations of Weyl's quantization prescriptions may
 be found.

\begin{Example} \ll{Lie}
Rieffel's quantization of the Lie-Poisson structure on a dual Lie
algebra
\end{Example}

A Lie group is a Lie groupoid with $Q=e$.  A left-invariant Haar
measure on $G$ provides a left Haar system; the ensuing convolution
algebra $C^*_r(G)$ is the usual reduced group algebra.  The Lie
algebroid is the Lie algebra. The Poisson structure on $\g_{\pm}^*$ is 
$\mp$ the well-known  Lie-Poisson structure \ci{MR,MT}.

No connection is needed to define the exponential map, and one has \be
\Exp^L(X)=\Exp^W(X)=\Exp(X),\ll{expWG} \ee where $X\in\g$ and
$\Exp:\g\raw G$ is the usual exponential map.  When $G$ is exponential
(in that $\Exp$ is a diffeomorphism), one may omit $\kp$ in
(\ref{qwfEW}). Taking the $+$ sign, the function $\qw(f)_+\in
C^*_r(G)$ is then given by \be
\qw(f)_+:\Exp(X)\raw\int_{\g^*}\frac{d^n\th}{(2\pi\hbar)^n}\,
e^{i\la\th,X \ra/\hbar}f(\th).\ll{defqhc} \ee This is Rieffel's
prescription \ci{Rie2}, who proved 
strictness of the quantization for nilpotent groups.  When
$G$ is compact one needs the cut-off function $\kp$, obtaining
another quantization already known to be strict before the
present paper and \ci{Ram} appeared; see \ci{LMP} or \ci{MT}.

\begin{Example} 
Weyl quantization on a gauge groupoid.
\end{Example}

The gauge groupoid $\SP\x_H\SP\daw Q$ of a smooth principal bundle
$\SP$ over a base $Q$ with structure group $H$ is defined by the
projections $\ta_s([x,y]_H)=\ta(y)$ and $\ta_t([x,y]_H)=\ta(x)$, and
the inclusion $\io(\ta(x))=[x,x]_H$. Accordingly, the multiplication
$[x,y]_H\cdot [x',y']_H$ is defined when $y$ and $x'$ lie in the same
fiber of $\SP$, in which case $[x',y']_H=[y,z]_H$ for some $z=y'h$,
$h\in H$. Then $[x,y]_H\cdot [y,z]_H=[x,z]_H$.  Finally, the inverse
is $[x,y]_H\inv=[y,x]_H$. See \ci{Mac}.

An $H$-invariant measure $\mu$ on $\SP$ which is locally Lebesgue
produces a left Haar system.  In general, each measurable section
$s:Q\raw \SP$ determines an isomorphism $C_r^*(\SP\tih\SP) \simeq
\B_0(L^2(Q))\ot C_r^*(H)$; this is a special
case of Thm.\ 3.1 in \cite{MRW} (also cf.\ \ci{MT}, Thm.\ 3.7.1).
When $H$ is compact one has
$C_r^*(\SP\tih\SP)\simeq \B_0(L^2(\SP))^H$, where $(L^2(\SP)$ is
defined with respect to some $H$-invariant locally Lebesgue measure on
$\SP$.

 The associated Lie algebroid $(T\SP)/H\aaw$ is defined by the obvious
projections (both inherited from the projection $\ta:\SP\raw Q$), and
the Lie bracket on $\Gm((T\SP)/H)$ obtained by identifying this space
with $\Gm(T\SP)^H$, and borrowing the commutator from $\Gm(T\SP)$;
cf.\ \ci{Mac}. The Poisson structure on $((T\SP)/H)^*=(T^*\SP)/H$ is
given by the restriction of the canonical Poisson bracket on
$\cin(T^*\SP)$ to $\cin(T^*\SP)^H$, under the isomorphism
$\cin((T^*\SP)/H)\simeq \cin(T^*\SP)^H$.

One chooses an $H$-invariant affine connection on $T\SP$, with
exponential map $\exp:T\SP\raw \SP$. This induces a connection on
$(T\SP)/H$, in terms of which \bea \Exp^L([X]_H) & = &
[\ta(X),\exp_{\ta(X)}(X)]_H ;\ll{expLPHP} \\ \Exp^W ([X]_H) & = &
[\exp_{\ta(X)}(-\half X),\exp_{\ta(X)}(\half X)]_H, \ll{expWPHP} \eea
where $\ta= \ta_{T\SP\raw\SP}$, and $[X]_H\in (T\SP)/H$ is the
equivalence class of $X\in T\SP$ under the $H$-action on $T\SP$.

In the Riemannian case, for compact $H$ the corresponding map 
$\qw(\cdot)_-$ is
simply the restriction of $\qw(\cdot)_-:\CPW(T^*\SP)\raw \B_0(L^2(\SP))$ as
defined in Example \ref{WM} to $\CPW(T^*\SP)^H$. Since $\qw$ is
invariant under isometries \ci{MT}, the image of $\CPW(T^*\SP)^H$ is
contained in $\B_0(L^2(\SP))^H$. The ensuing quantization of
$(T^*\SP)/H$ was already known to be strict; see
\ci{NPL1,MT}. Physically, this example describes the quantization of a
nonabelian charged particle moving in a gravitational as well as a
Yang-Mills field.

\begin{Example}
Transformation group \ca s
\end{Example}

Let a Lie group $G$ act smoothly on a set $Q$. The transformation
group\-oid $G\x Q\daw Q$ is defined by the operations
$\ta_s(x,q)=x\inv q$ and $\ta_t(x,q)=q$, so that the product
$(x,q)\cdot (y,q')$ is defined when $q'=x\inv q$. Then $(x,q)\cdot
(y,x\inv q)=(x y,q)$. The inclusion is $\io(q)=(e,q)$, and for the
inverse one has $(x,q)\inv=(x\inv,x\inv q)$.

Each left-invariant Haar measure $dx$ on $G$ leads to a left Haar
 system. The corresponding reduced groupoid \ca\ is the usual reduced
 transformation group \ca\ $C_r^*(G,Q)$, cf.\ \ci{Ren}.

The Lie algebroid $\g\x Q\aaw$ is a trivial bundle over $Q$, with
 anchor $\ta_a(X,q)=-\xi_X(q)$ (the fundamental vector field on $Q$
 defined by $X\in\g$). Identifying sections of $\g\x Q$ with
 $\g$-valued functions $X(\cdot)$ on $Q$, the Lie bracket on $\Gm(\g\x
 Q)$ is \be [X,Y]_{\g\x Q}(q)=[X(q),Y(q)]_{\g}+\xi_Y X(q)-\xi_X
 Y(q). \ll{LBactioloid} \ee The associated Poisson bracket coincides
 with the semi-direct product bracket defined in \ci{KM}.

The trivial connection on $\g\x Q\raw Q$ yields \bea \Exp^L(X,q) & = &
(\Exp(X),q) ;\ll{expLAG}\\ \Exp^W(X,q) & = & (\Exp(X),\Exp(\half
X)q).\ll{expWAG} \eea The cutoff $\kp$ in (\ref{qwfEW}) is independent
of $q$, and coincides with the function appearing in Example
\ref{Lie}.  For small enough $\hbar$ a function $f\in \CPW(\g^*\x Q)$
is then quantized by \be \qw(f)_{\pm}:(\Exp(X),q)
\raw\int_{\g^*}\frac{d^n\th}{(2\pi\hbar)^n}\,
e^{i\la\th,X\ra/\hbar}f(\pm\th,\Exp(-\half X)q). \ll{qwongact} \ee
When $G=\R^n$ and $Q$ has a $G$-invariant measure, the map $f\raw
\qw(f)_{\pm}$ is equivalent to the deformation quantization considered
by Rieffel \ci{Rie1}, who already proved that it is strict (also cf.\
\ci{MT}).

\end{document}